# Analysis of Femoral Components of Cemented Total Hip- Arthroplasty

Shantanu Singh[a,1], A.P. Harsha[a]

[a]Department of Mechanical Engineering, Indian Institute of Technology (Banaras Hindu University), Varanasi 221005, Uttar Pradesh, India.
E-mail: shantanu.singh.mec09@iitbhu.ac.in

**ABSTRACT**

In cemented Total Hip Arthroplasty (THA), material chosen for femoral stem and cross section of stem itself, proved to be critical parameters for, stress distribution in the femoral components, interfacial stresses and micro movements. Titanium alloy ($Ti_6Al_4V$), when used as a material for femoral stem, recorded large displacement as compared to Chromium alloy (CoCrMo) stems. This large displacement in case of $Ti_6Al_4V$ caused the stem to bend inside the cement mantle, thus destroying it. Thus, CoCrMo proved to be a better choice of material for stem, in cemented THA. Failure in THA may occur at cement-stem or cement-bone interface, thus interfacial stresses and micro movements were analysed in the present study. Comparison between trapezium and circular cross section showed that, femoral stem with trapezium cross section underwent lesser amount of sliding and debonding, at both interfaces, as compared to circular cross section. Moreover, trapezium cross section also generated lower peak stresses in femoral stem and cortical femur. The present study also took into account, effect of diameter of femur head. A 36mm diameter femur head generated lower peak stress in acetabulum liner and arrested dislocation. Metallic femur head was coupled with a liner made of cross linked polyethylene (XLPE). This material experiences almost negligible wear when compared to typical metallic and polyethylene liners, and unlike metallic liner, it is non-carcinogenic.

## 1. INTRODUCTION

Total Hip Replacement (THR) is an excellent alternative for patients suffering from arthritis or hip fracture. Millions of people undergo total hip replacement every year [1]. THR results in reduced hip pain, increased mobility and an overall better quality of life for the patient. Depending upon patient, Total Hip Arthroplasty (THA) may be cemented or cementless. Cemented hip stems have the advantage of less rehabilitation time as the patient is able to walk without support soon after surgery. Generally, it is preferred for old patients, while cementless THA is preferred for young patients with high physical activity [2]. Increased physical activities also demand THA with large femoral head diameter to prevent dislocation [3, 4]. Occurrences of dislocation, ranges from 5% to 25% in revision surgeries, and 2 to 11% in primary surgeries [5]. Metal on metal bearing allows for large diameter femoral head and may seem to be the ideal choice for bearing combination, but it suffers from a serious drawback. During motion of the hip joint, wear of metal occurs, which results in an increased concentration of metallic ions in blood. Therefore, this type of implant is suspected of having carcinogenic effects. Also, in case of polyethylene on metal bearings, large diameter femur head produces significant, Ultra high molecular weight polyethylene (UHMWPE) wear debris. Thus, present study considers combination of metallic femur head and liner made of cross-linked polyethylene (XLPE), which relatively undergoes negligible wear [1].

In present study, analysis was carried out with femoral head diameter of 28 mm, 32mm and 36 mm. Implants with circular and trapezium cross sectional stems were considered and stress distribution was analysed at interfaces and femoral components. In many cases of failure of cemented hip joint, delamination of femoral stem from Poly methyl methacrylate (PMMA) bone cement occurs. This debonding is also associated with the torsional movements of implant. Failure may also occur at cement bone interface, thus it is important to analyze interfacial stresses and micro movements.

In order to control micro motion, the design of hip prosthesis should be such that it experiences an even distribution of stress avoiding any possible stress concentration, which may cause failure of the joint. Moreover, the bone density and shape around the implant changes due to various loads on the prosthesis [6, 12]. As bone density changes due to uneven distribution of stress, it alters fixation of the implant. In present study, both cement-stem and cement–bone interfaces were considered to have frictional contact. Optimal values of coefficient of friction were taken so as to minimize stresses and micro movement at both the interfaces [13]. Unlike many of the previous studies, where interfaces were often considered to be fully bonded, present study mimics the actual THA condition.

## 2. METHOD OF ANALYSIS

### 2.1. Material properties

Six different materials were considered in the present investigation to study the performance of THA. Cobalt Chromium alloy (CoCrMo) and Titanium alloy ($Ti_6Al_4V$) were the materials considered for femoral stem. Other materials considered were cross-linked polyethylene (XLPE) for the liner, Poly methyl methacrylate (PMMA) as bone cement, cancellous bone and cortical bone. All these materials have excellent biocompatibility [5, 6]. The materials considered were assumed to be homogeneous, isotropic and linear elastic solids. **Table 1** lists the different material properties.

**Table 1. Material properties used in FE model**

| Material | Elastic Modulus (GPa) | Poisson's Ratio | Yield Strength (GPa) | Ultimate Tensile Strength (GPa) | References |
|---|---|---|---|---|---|
| CoCrMo | 230 | 0.3 | 0.455 | 0.65 | [15,17] |
| $Ti_6Al_4V$ | 110 | 0.3 | 0.795 | 0.86 | [14,16] |
| Cross Linked Polyethylene | 1.4 | 0.3 | 0.023 | 0.044 | [15] |
| Bone Cement | 2.5 | 0.38 | - | 0.0353 | [5,16] |
| Cortical Bone | 17 | 0.29 | - | - | [18] |
| Cancellous Bone | 0.52 | 0.29 | - | - | [5] |

### 2.2. Finite element (FE) model

The three-dimensional geometry of femur was reconstructed using quantitative computer tomography (CT), while computer-aided design (CAD) model of other femoral components including the femoral stem were developed in CATIA V5 software. After assembling the full model of THA, ANSYS 14.0 Release was used to develop the FE model of the hip joint. **Figure 1** shows 3-D model of the artificial hip joint.

Ramaniraka et al. [13] reported that slipping and debonding are minimal for cement thickness range of, 3mm to 5 mm, thus average thickness of bone cement was taken as 4 mm. The femoral stem was bonded to bone by PMMA bone cement, while femoral head forms a ball and socket joint in a XLPE liner. In present study, femoral stems made up of CoCrMo and $Ti_6Al_4V$ were considered respectively.

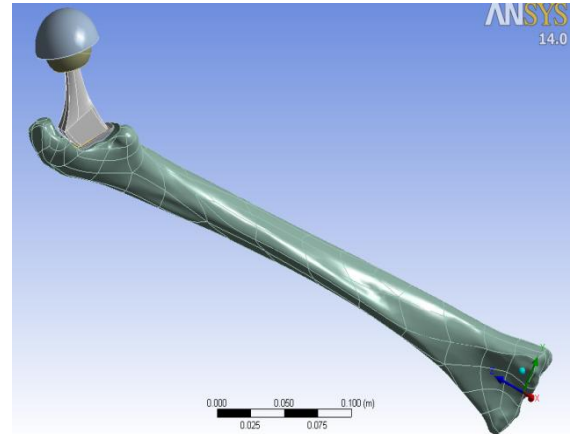

Figure 1. 3-D model of artificial hip joint

**Table 2** and **3** give details of FE model considered in present study.

**Table 2. FE model details**

| Object | Femoral stem | XLPE liner | Femoral head | Cancellous Bone | PMMA bone cement | Cortical bone |
|---|---|---|---|---|---|---|
| Mesh | Hex-Dominant (SOLID 186, SOLID 187 element) | | | Automatic | | |
| Element size | 2 mm | | | 10 mm | | |

**Table 3. Contact details**

| Femur head-Stem | Cancellous-Cortical | PMMA-Cancellous | Femur head-XLPE | Stem-PMMA |
|---|---|---|---|---|
| Bonded | Frictional contact (µ=1) | Frictional contact (µ=1) | Frictional contact (µ=0.2) | Frictional contact (µ=0.4) |

Optimum values of coefficient of friction, 0.4 for cement-stem and 1.0 for cement-bone interfaces, were considered respectively [13].

**Figure 2(a)** and **2(b)** shows the hip profile with trapezium and circular cross section, used in present study [6]. Fillet of 2 mm was provided in trapezoidal stem to avoid stress concentration.

Both SOLID 186 and SOLID 187 elements were used for meshing. SOLID 186 elements are 20-noded while SOLID 187 are 10-noded. They are higher order 3-D elements that exhibit quadratic displacement behaviour, with each node of the elements having three degrees of freedom. These attributes make the elements suitable for analysis. Different mesh capacities were employed in order to ensure convergence of the solution, followed by mesh optimization.

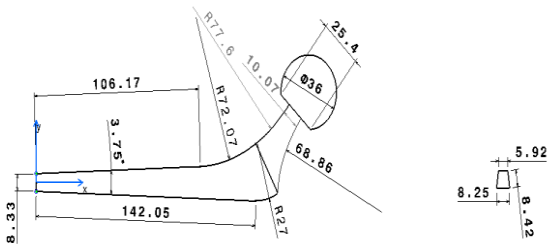

Figure 2(a). Hip profile with distal cross-section as trapezium (Dimensions in mm)

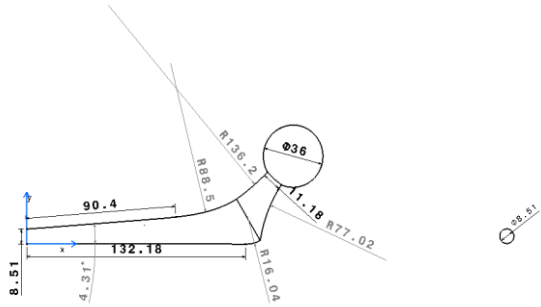

Figure 2(b). Hip profile with distal cross-section as circle (Dimensions in mm)

### 2.3. Application of loads and constraints

In present study, loading conditions corresponded to single-limb stance in gait cycle. The load on femur head was simulated with a force which was three times the body weight of a patient (60 years old weighing 600 N) [13]. Muscle forces of gluteus minimus, medius and maximus, and iliopsoas were also taken in account. The proximal end of femur bone was partially fixed and distal end was fully fixed. Details of loads and constraints were as shown in **Figure 3**.

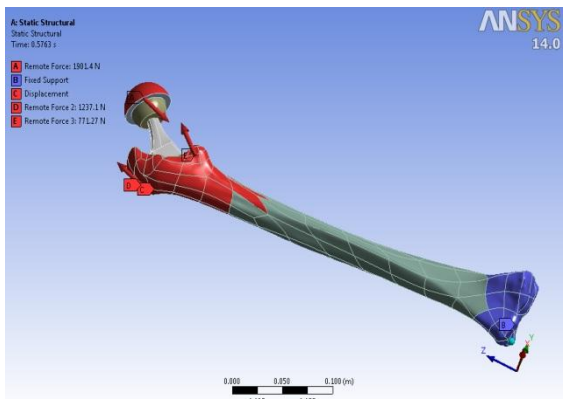

Figure 3. Boundary conditions and loads on THA

Spatial reference system was defined as follows: lateral-medial direction is denoted as x, anterior-posterior direction as y and vertical direction as z (up positive) [13]. **Table 4** lists the values of different forces acting on THA.

Table 4. Forces acting on THA

|  | $F_X$ [N] | $F_Y$ [N] | $F_Z$ [N] |
|---|---|---|---|
| Femur head | -320 | 448 | -1820 |
| Abductors | 430 | 0 | 1160 |
| Iliopsoas | 75 | 525 | 560 |

## 3. RESULTS AND DISCUSSIONS

### 3.1. Effect of material of stem

Two different materials were used for the femoral stem viz. CoCrMo and $Ti_6Al_4V$. It was observed that $Ti_6Al_4V$ femoral stem underwent large deformation within the cement mantle as compared to CoCrMo stem. This happened because, $Ti_6Al_4V$ had relatively lower stiffness than CoCrMo, and thus it experienced larger deformation while developing lower peak stress.

The maximum von-Mises stress developed in $Ti_6Al_4V$ stem was found to be 18.66 MPa while peak stress value for CoCrMo implant was 20.1 MPa.

Focusing towards stress distribution in cortical bone, it was found that, $Ti_6Al_4V$ stem resulted in a peak von-Mises stress of 6.21 MPa, while CoCrMo stem caused a peak stress of 5.44 MPa in cortical bone. Stress distribution patterns showed that lower stiffness material loads the proximal end, while stiffer materials load the distal end, an observation which was concurrent with previously reported results [19]. Maximum deformation for $Ti_6Al_4V$ stem was found to be 85.33 microns as compared to 46.10 microns for CoCrMo. Results showed that $Ti_6Al_4V$ stem bend up to an extent within the cement mantle so as to damage it, thus making CoCrMo the preferred material for cemented hip joints.

A high rate of dislocation has been quoted in case of hip replacement with smaller ball (femur head) size [4]. Recently, many active and young patients are undergoing hip replacements. They expect THA which could ensure high amount of physical activity, for their active lifestyles. Literatures suggest that large diameter femur head decreases the risk of dislocations even in complex revision surgery [4]. Modern solutions have been reported to this problem in form of cross-linked polyethylene liners and ceramic liners, which allow for large diameter femur head with relatively, very low wear rates. Although long term results are not available for these novel alternatives, but data of gait analysis proves that, larger diameter balls showed better gait pattern. [20,21]. In present study, analysis was carried out with femur head of diameter 28 mm, 32 mm and 36 mm. **Figure 4** shows the variation of peak von-Mises stress developed in the liner, against diameter of femur head.

Increasing diameter of femur head, decreased peak stress developed in liner without any significant effect on other femoral components. Owing to the above results, further discussions are based on implants made of CoCrMo with a femur head of diameter 36mm.

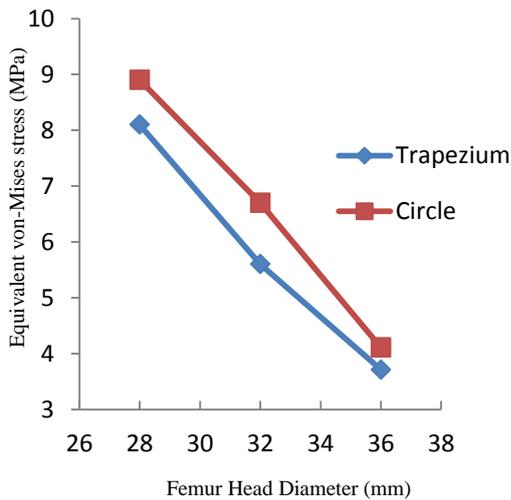

Figure 4. Variation of peak stress value in liner with respect to femur head diameter

### 3.2. Effect of cross-section of stem

The cross-section of femoral stem also had major effect on the stress distribution in THA. In present study, we considered stems of circular and trapezium cross-section, and compared them for stresses developed in bone, stem and interfaces as well.

Peak von-Mises stress value in trapezoidal stem was 20 MPa, as compared to 41.11 MPa in case of stem with circular cross-section, under same loading conditions. As shown in **Figure 5** and **6**, peak stress in trapezoidal stem occurred in the posterior region of central portion, whereas in circular stem, peak stress occurred in the neck region of stem.

Peak von-Mises stress value in cortical bone was 1.83 MPa, when the stem had trapezoidal cross section, as compared to value of 3.06 MPa in case of circular cross-sectional stem.

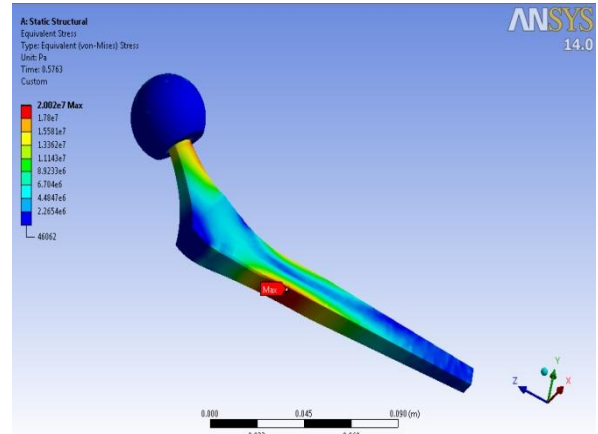

Figure 5. Stress distribution in CoCrMo femoral stem with trapezium cross section

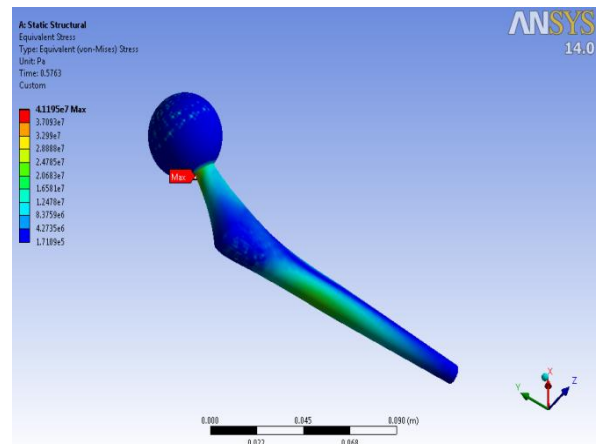

Figure 6. Stress distribution in CoCrMo femoral stem with circular cross section

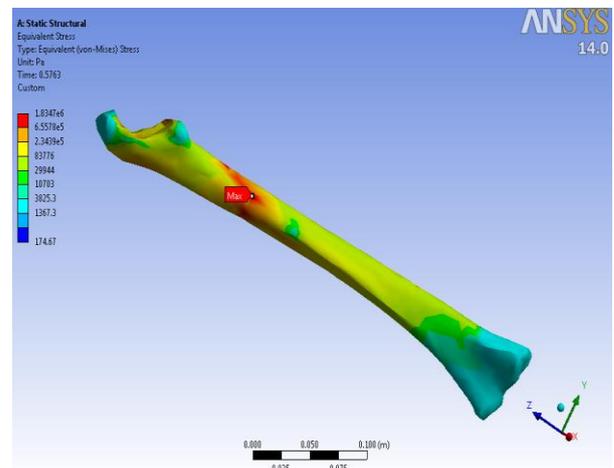

Figure 7. Stress distribution in Cortical bone for THA with trapezoidal cross sectional stem

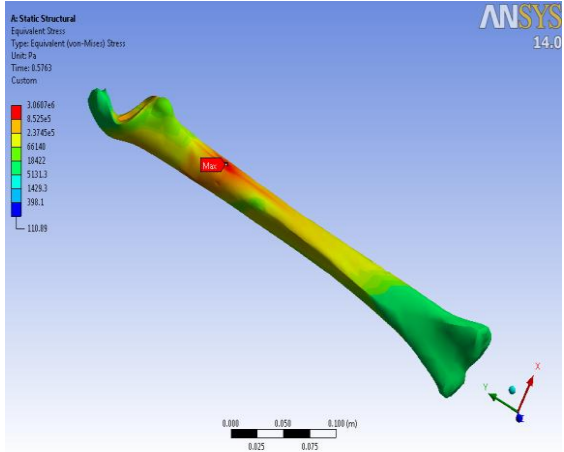

Figure 8. Stress distribution in Cortical bone for THA with circular cross sectional stem

From **Figure 7** and **8** it was observed that peak stress values in both cases occurred at the anterolateral proximal cortical femur, however trapezoidal stem resulted in a better stress distribution in distal and proximal femur.

**Figure 9** lists a comparative of peak stress values for different femoral components, between trapezoidal and circular stem.

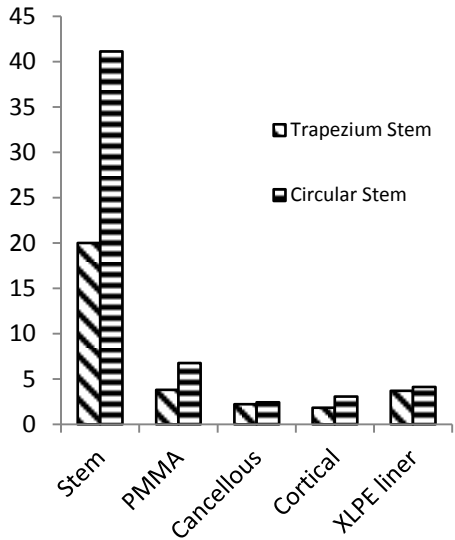

Figure 9. von- Mises stress histogram for different femoral components

**Table 5** and **6** lists the peak stress and micro movement at cement-stem and cement-bone interface.

**Table 5. Stress and micro movement for cement-stem interface**

|  | Trapezoid | Circular |
|---|---|---|
| Pressure (MPa) | 6.51 | 3.88 |
| Frictional stress (MPa) | 2.30 | 0.76 |
| Sliding (μm) | 51.74 | 93.60 |
| Debonding(μm) | 39.06 | 53.11 |

**Table 6. Stress and micro movement for cement-bone interface**

|  | Trapezoid | Circular |
|---|---|---|
| Pressure (MPa) | 2.9 | 2.1 |
| Frictional stress (MPa) | 0.7 | 0.8 |
| Sliding (μm) | 15 | 28 |
| Debonding(μm) | 103 | 131.07 |

Debonding was observed mainly in proximal and distal ends of cement-stem interface, while in case of cement-bone interface, debonding occurred in proximal and medial region. It was observed that slipping and debonding in case of trapezoidal stem was less than that in circular stem, both for cement-stem and cement-bone interface. Pressure developed at both interfaces was lesser in case of circular stem and in all cases pressure was well below 65 MPa, which is allowable strength for PMMA [16]. **Table 7** lists a comparative between present study and previously reported work. In this study, both the trapezoidal design of stem and large diameter femoral head used reduced dislocation to a large extent. Moreover, trapezoidal stems have better torsional stability than circular stems [22]. Care should be taken not to contaminate the stem surface or cement surface during arthroplasty, as any contamination leads to significant reduction in shear strength of interface [23].

**Table 7. Comparison between peak stresses and micro movements at cement-stem interface**

|  | Present study | Ramaniraka et al.[13] |
|---|---|---|
| Pressure (MPa) | 6.51 | 6.58 |
| Frictional stress (MPa) | 2.30 | 2.63 |
| Sliding (μm) | 51.74 | 109 |
| Debonding(μm) | 39.06 | 63.1 |

## 4. CONCLUSION

Finite element analysis results showed that femoral stem with trapezium cross-section was better than stem with circular cross-section. Trapezoidal stem resulted in lower interfacial micro movements, and developed lower peak stresses in different femoral components. It was observed that CoCrMo was the preferred material for cemented THA. In future Fatigue loading can also be taken into account and dynamic analysis could be performed over whole gait cycle of walking. Moreover, due to computational limitations, different materials used in the present study, which are actually anisotropic, have been assumed to be isotropic in nature. Long term clinical followup of patients having THA with large femur head diameter should be done to, ascertain the effect of large femur head on dislocation. For a more optimized design a combination of different cross sections needs to be studied which may more likely result into better stress distribution.

**Acknowledgements**

We would like to acknowledge the support of King George Medical University, Lucknow, India for providing us with computer tomography (CT) images of femur.

## REFERENCES

[1] Harsha, A.P. and Joyce, T.J. (2013) Comparative wear tests of ultra-high molecular weight polyethylene and cross-linked polyethylene. Journal of Engineering in Medicine, **227(5)**, 600-608. http://dx.doi.org/10.1177/0954411913479528

[2] Hailer, N.P., Garellick, G., and Kärrholm, J. (2010) Uncemented and cemented primary total hip arthroplasty in the Swedish Hip Arthroplasty Register, Evaluation of 170,413 operations. Acta Orthop., **81(1)**, 34–41. http://dx.doi.org/10.3109/17453671003685400

[3] MacDonald, S.J. (2004) Metal-on-Metal Total Hip Arthroplasty. Clinical orthopaedics and related research, **429**, 86–93. http://dx.doi.org/10.1097/01.blo.0000150309.48474.8b

[4] Fender, D., Harper, W.M. and Gregg P.J. (1999) Outcome of Charnley total hip replacement across a single health region in England: the results at five years from a regional hip register. Journal of Bone & Joint Surgery (Br), **81(4)**, 577-81. http://dx.doi.org/10.1302/0301-620X.81B4.9859

[5] Darwish, S.M. and Al-Samhan, A.M. (2009) Optimization of Artificial Hip Joint Parameters. Mat.-wiss. u. Werkstofftech, **40(3)**, 218-223. http://dx.doi.org/10.1002/mawe.200900430

[6] Sabatini, A.L. and Goswami, T. (2008) Hip implants VII: Finite element analysis and optimization of cross-sections. Materials and Design, **29,** 1438–1446. http://dx.doi.org/10.1016/j.matdes.2007.09.002

[7] Gebauer, D., Refior, H.J., and Haake, M. (1989) Micro motions in the primary fixation of cementless femoral stem prostheses. Archives of Orthopaedic and Trauma Surgery, **108(5)**, 300-307. http://dx.doi.org/10.1007/BF00932320

[8] Viceconti, M., Monti, L., Muccini, R., Bernakiewicz, M. and Toni, A. (2001) Even a thin layer of soft tissue may compromise the primary stability of cementless hip stems. Clinical Biomechanics, (Bristol, Avon), **16 (9)**, 765-775. http://dx.doi.org/10.1016/S0268-0033(01)00052-3

[9] Engh, C.A., O'Connor, D., Jasty, M., McGovern, T.F., Bobyn, J.D. and Harris, W.H. (1992) Quantification of implant micro motion, strain shielding and bone resorption with porous-coated anatomic medullary locking prosthesis. Clinical Orthopaedics and Related Research, **285**, 13–29. DOI 10.1097/00003086-199212000-00005

[10] Søballe, K., Hansen, E.S., Brockstedt-Rasmussen, H. and Bunger, C. (1993) Hydroxyapatite coating converts fibrous tissue to bone around loaded implants. The Journal of Bone & Joint Surgery, **75(2)**, 270–278.

[11] An, Y.H. and Draughn, R.A. (2000) Mechanical Testing of Bone and the Bone–Implant Interface. CRC Press.

[12] Carter, D.R., Orr, T.E. and Fyhrie, D.P., (1989) Relationship between loading history and femoral cancellous bone architecture. Journal of Biomechanics, **22(3)**, 231–244. http://dx.doi.org/10.1016/0021-9290(89)90091-2

[13] Ramaniraka, N.A., Rakotomanana, L.R. and Leyvraz, P.F. (2000) The fixation of cemented femoral component. Journal of Bone & Joint Surgery (Br), **82-B,** 297-303. http://dx.doi.org/10.1302/0301-620X.82B2.9613

[14] Pyburn, E. and Goswami, T. (2004) Finite element analysis of femoral components paper III – hip joints. Materials and Design, **25(8)**, 705–713. http://dx.doi.org/10.1016/j.matdes.2004.01.009

[15] Bronzino, J.D. (2006) Biomedical Engineering Fundamentals. CRC Press .

[16] Bronzino, J.D. (2000) The Biomedical Engineering Handbook, volume 1. CRC Press.

[17] Kayabasi, O. and Erzincanli, F. (2006) Finite element modelling and analysis of a new cemented hip prosthesis. Advances in Engineering Software, **37(7)**, 477–483. http://dx.doi.org/10.1016/j.advengsoft.2005.09.003

[18] Monif, M.M. (2012) Finite element study on the predicted equivalent stresses in the artificial hip joint. Journal of Biomedical Science and Engineering, **5**, 43-51. http://dx.doi.org/10.4236/jbise.2012.52007

[19] Huiskes, R. (1991) Chapter 9, Biomechanics of Artificial-joint Fixation. In: Mow, V.C. and Hayes, W.C., Ed., Basic orthopaedic Biomechanics. Raven Press, Ltd., New York, 375-442.

[20] Nantel, J., Termoz, N., Centomo, H., Lavigne, M., Vendittoli, P.A. and Prince F. (2008) Postural balance during quiet standing in patients with total hip arthroplasty and surface replacement arthroplasty. Clinical Biomech

(Bristol, Avon), **23(4)**, 402-407. http://dx.doi.org/10.1016/j.clinbiomech.2007.10.011

[21] Nantel, J., Termoz, N., Ganapathi, M., Vendittoli, P.A., Lavigne, M. and Prince, F. (2009) Postural balance during quiet standing in patients with total hip arthroplasty with large diameter femoral head and surface replacement arthroplasty. Archives of physical medicine and rehabilitation, **90(9)**, 1607-1612. http://dx.doi.org/10.1016/j.apmr.2009.01.033

[22] Keggi, K.J., Keggi J., Kennon, R., Keppler, L., Turnbull, A. and McTighe, T. (2009) Proximal Modular Stem Design "Dual Press" With a Dual-Tapered "K2" Trapezoid Stems. Joint Implant Surgery & Research Foundation, Chagrin Falls, Ohio, USA.

[23] Stone, M.H., Wilkinson, R. and Stother I.G. (1989) Some factors affecting the strength of the cement-metal interface. The Journal of Bone & Joint Surgery, **71(2)**, 217-221. DOI 0301-620X/89/2035